\documentstyle[preprint,aps]{revtex}

\title{ The $c$~-term of the TM 3-body Force: to be or not to 
be\thanks{ Dedicated to Prof. Dr. Walter Gl\"ockle on the occasion of his 
birthday}
}
\author{H. Kamada$^\S$\thanks{{\it Alternative address:}
Institut f\"ur Strahlen- und Kernphysik der Universit\"at Bonn,
Nussallee 14-16, D-53115 Bonn,
    Germany  }  
\thanks{{\it E-mail address:}
kamada@hadron.tp2.ruhr-uni-bochum.de}
D. H\"uber$^{\S~\ddag}$
\thanks{{\it E-mail address:}
hueber@paths.lanl.gov}
 and A. Nogga$^\S$
\thanks{{\it E-mail address:}
andreasm@hadron.tp2.ruhr-uni-bochum.de}}
\address{$^\S$ Institut f\"ur Theoretische Physik II,
Ruhr-Universit\"at Bochum, D-44780 Bochum, Germany \\
$^\ddag$  Theoretical Division, M.S. B283, Los Alamos National Laboratory, 
Los Alamos, New Mexico 87545, USA}

\begin{document}

\maketitle

\begin{abstract}
In Faddeev calculations of $^3$H 
we study the dependence of the binding energy on the three nucleon force.
We adopt the $2\pi$-exchange
Tucson-Melbourne three-nucleon force 
and investigate phenomenologically the dependence on the  strength
of the individual  
 three-body force operators
($a$-, $b$-, $c$- and $d$-terms). 
While the $a$-term provides a tiny contribution the 
$b$- and $d$-terms  are important to gain the experimental binding energy. 
We find two solutions for the $c$-term, one around the value used in the Tucson-Melbourne 
model and a new one close to zero, which  supports the recent recommendation of chiral 
perturbation theory that  the short-range  $c$-term should be 
dropped. 
\end{abstract}

\section{Introduction}
It is one of the great dreams in  the field of  few-nucleon physics 
to find a quantitative correct
and theoretically reasonable three nucleon force(3NF).
In the past many 3NF models were developed\cite{Fujita,Tuscon,Coon81,Brazil,TEXAS,Ruhr,Urbana}.
An  especially prominent one is the meson-theoretical 3NF, for  instance in the form of the Tucson-Melbourne 
model (TM)\cite{Tuscon,Coon81}. 
The reason for studying 3NFs is the existence of  
disagreements between the 3N data and the theoretical predictions with 
NN forces only. 
First of all 
the theoretical binding energy of  
$^3$H lacks about 500-800keV in relation to the experimental value of 8.48MeV 
using recent realistic potentials (e.g. CD-Bonn\cite{cdbonn}, 
AV18\cite{av18}, Nijmegen 93,
Nijmegen I, II\cite{nijm}).  
These potentials  describe all 2N observables to a degree of accuracy of $\chi ^2 / N_{data} \sim$1. 
In the low energy three nucleon continuum we have 
demonstrated\cite{Gloeckle96} that
most of the observables agree well with the data using nucleon-nucleon (NN) forces only,
however, there are exceptions.  
Some of them are well known as the  ``$A_y$ puzzle''\cite{Koike,Witala3,Witala,Hueber,Kievsky}.
In the high energy region the theoretical predictions differ visibly from the data if one 
only takes NN forces into account. 
The 
``Sagara discrepancy''\cite{Sagara,Koike2,Witala2,Sakamoto,Nemoto} is an example of this. 
These problems  definitely require 
a new Hamiltonian in the realm  of the three nucleon system.
Moreover, the $A_y$-puzzle requires not only a $2\pi$-exchange 3NF to become
explained\cite{Gloeckle96,Kievsky} but other 3NF mechanisms as well.

Beside these low-energy discrepancies in the 3N continuum there are also
discrepancies at higher energies. This can be expected naively due to the
shorter range nature of the 3NF in comparison to the NN force. Recently it
became possible to explain discrepancies between experiment and predictions
using NN forces only for the neutron-deuteron (nd) total cross section 
\cite{Total} and the nd elastic differential cross section \cite{Diff} with
the $2\pi$-exchange TM 3NF.

From the point of view of  chiral perturbation, this special category of the
TM 3NF should be modified\cite{Friar99}.
Chiral perturbation theory has been successfully applied  
in the $\pi N$ system \cite{piNchiral,piNchiral2} 
 and it is already playing an important role in its application to the NN system as 
well\cite{NNchiral,NNchiral2,NNchiral3}
In \cite{Friar99} it recommended that the pion range - short range part
of the $c$-term of the TM $2\pi$-exchange 3NF should be dropped, based on
arguments from chiral perturbation theory. In doing this the pion range -
short range part of the $c$-term remains, which is of the same type than the
$a$-term. This leads to a redefinition of $a$.
The such modified TM 3NF,called TM', has essentially the same effects on 
continuum\cite{Hueber2} than the original TM 3NF.
Much remains to be investigated in the relation between NN and 3NF's.

In the next section we present calculations for the triton binding energy based 
on variations of the values of the strength parameters in the
TM 3NF, individually and combined.
The summary and  the outlook are give in Section 3.

\section{Variations of the Tucson-Melbourne 3NF and their Triton Binding Energies}

The TM force has the operator form:

\begin{eqnarray}
V_{TM}^{(3)} ={ 1 \over{ (2\pi)^6 }} { {g^2 _{\pi NN} }\over { 4 m^2 }}
{ { F^2_{\pi NN} (\vec q^2 ) F^2_{\pi NN}(\vec {q'}^2)
\vec \sigma_ 1 \cdot \vec q 
\vec \sigma _2 \cdot \vec {q'} } \over { ( \vec q ^2 + m_\pi^2  ) ( \vec 
{q '} ^2 + m_\pi  ^2 ) }} 
[ O^{\alpha \beta } \tau_\alpha \tau_\beta ],
\end{eqnarray}
\begin{eqnarray}
O^{\alpha \beta } =\delta ^{\alpha \beta} [ 
a + b \vec q \cdot \vec {q'} + c (\vec q ^2 + \vec {q'} ^2 )]
-d ( \tau _3 ^\gamma \epsilon ^{ \alpha \beta \gamma } \vec \sigma _3 
\cdot \vec q \times \vec {q'} ).
\label{operator}
\end{eqnarray}
where $m_\pi$, $m$, $g _{\pi NN}$ and  $ F_{\pi NN} (\vec q^2 )$ are 
the pion mass, the nucleon mass, the $\pi NN$ coupling constant and the vertex function ,
respectively. The superscript (3) denotes that this expression is only one of three 
cyclically permuted parts of the total TM 3NF. There are four parameters
($a$, $b$, $c$ and $d$ ) which are chosen according to certain physical concepts\cite{Tuscon,Coon81}.
For practical calculations one needs to introduce the vertex function which 
is normally chosen as 
\begin{eqnarray}
F_{\pi NN}(\vec q ^2) = { { \Lambda^2 - m_\pi ^2 } \over { \Lambda ^2 + 
\vec q ^2 }}
\end{eqnarray}
The triton binding energy turns out to be strongly dependent  
on the cut-off parameter $\Lambda$.
In a phenomenological approach it can be used as a fit parameter to adjust the 
triton binding energy and this separately for each NN 
potential\cite{Stadler,Nogga}.
Using these cut-off parameters we calculated the polarization  transfer parameter 
$K_y^{y'}$ in the three-body continuum. While the individual pure NN force predictions are 
different they essentially coincide 
if those individually adjusted 3NFs where included and that prediction agrees rather well with the data\cite{Hempen98}.
This is one example out of several where scaling with  the triton binding energy exists for 3N continuum observables. In those studies we kept the original TM parameters ($a$, $b$, $c$ and $d$) and only varied the form-factor cut-off parameter$\Lambda$. 


Now we want to go one step further and study phenomenologically the dependence of the triton binding energy on the individual terms in the TM 3NF operators connected to the $a$-, $b$-,$c$- and  $d$-term. 
Like in \cite{Nogga} we solve 
the Faddeev equation rigorously including  the 3NF.
We choose  CD-Bonn  as the NN interaction. 
The original parameters\cite{Coon81} of the TM model are given in Table 2.1.  
\begin{table}
\caption{ Parameters for the original 3NF.
\label{table1}}
\begin{tabular}{lccccccc}
$a $  [$m_\pi^{-1}$]  &  $b$ [$m_\pi^{-3}$] & $c$ [$m_\pi^{-3}$] &
  $d$   [$m_\pi^{-3}$]
  &  $m_\pi$ [MeV] & $m$ [MeV]  &$g ^2 _{\pi NN}$ & $\Lambda $ [$m_\pi$]   \\
  \hline
 1.13 & -2.58 & 1.00  & -0.753   & 139.6  &938.926 &179.7  &  4.856  \\
  \end{tabular}
  \end{table}
The cut-off parameter $\Lambda$ is not the original one but adjusted to reproduce
 the triton binding energy together with CD-Bonn. 
We multiply each one of the parameters $a$, $b$, $c$ and $d$ by a factor $X$
($ 0 \le X \le 1.5$ ), one after the other:    
\begin{eqnarray}
\pmatrix{
a \cr 
b \cr 
c\cr
d\cr} \longrightarrow \cases{ 
\pmatrix{ a X\cr
b \cr
c\cr
d\cr} & case a \cr
\pmatrix{ a  \cr
b X \cr
c\cr
d\cr} & case b \cr
\pmatrix{ a \cr
b \cr
c X\cr
d\cr} & case c \cr
\pmatrix{ a \cr
b \cr
c\cr
d X \cr} & case d \cr }
\end{eqnarray} 
and determine  the 3N binding energy for these four cases.
The results are  shown in Fig.~1. 
\begin{figure}[hbt]
\input{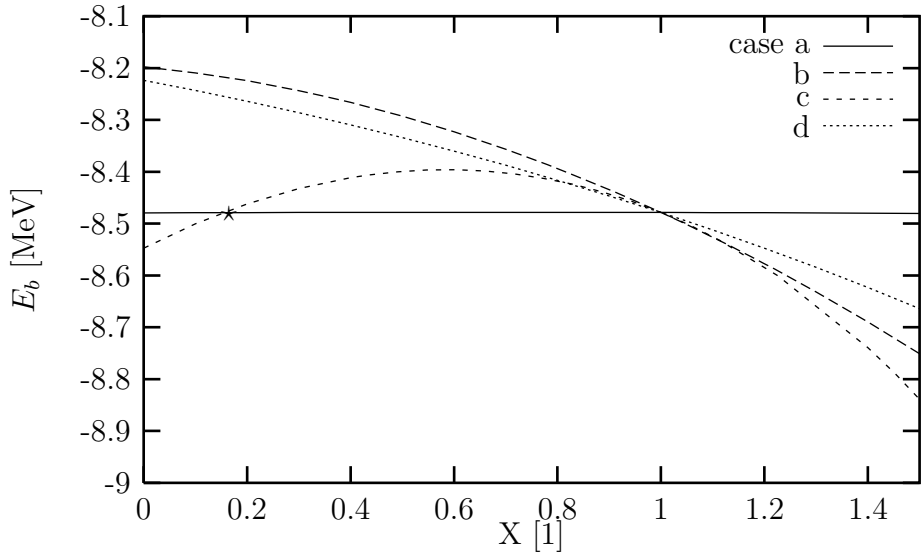}
\caption{The binding energy of the triton. The star ``$\star$'' is a new solution for the $c$-parameter.  }
\label{fig1}
\end{figure}

We see that the parameter $a$ contributes negligibly to the 3N bound state 
and its presence or absence is unimportant.
This explains why the prediction for the triton binding energy for the TM and TM' 3NFs are
close to each other\cite{Hueber2}.
The $b$- and $d$-terms however are important.  The binding energy increases monotonically with their strength.
Interestingly, the behaviour of the $c$-term is such that there are two solutions 
which lead to the experimental value.
We find  0.150 as the new  solution (see ``$\star$'' in Fig.1).  

Now, with the exception of $a$ we let all parameters float.
The parameter $a$ is kept at its  original value 1.13. We search for the sets  ($b$, $c$, $d$) 
which fulfil the condition to produce the experimental binding energy.
Thus we have now three independent  variables $X$: 
\begin{eqnarray}
\pmatrix{
a \cr
b \cr
c\cr
d\cr} \longrightarrow 
\pmatrix{ a \cr
b X_b  \cr
c X_c \cr
d X_d \cr}  
\end{eqnarray}
\begin{figure}[hbt]
\input{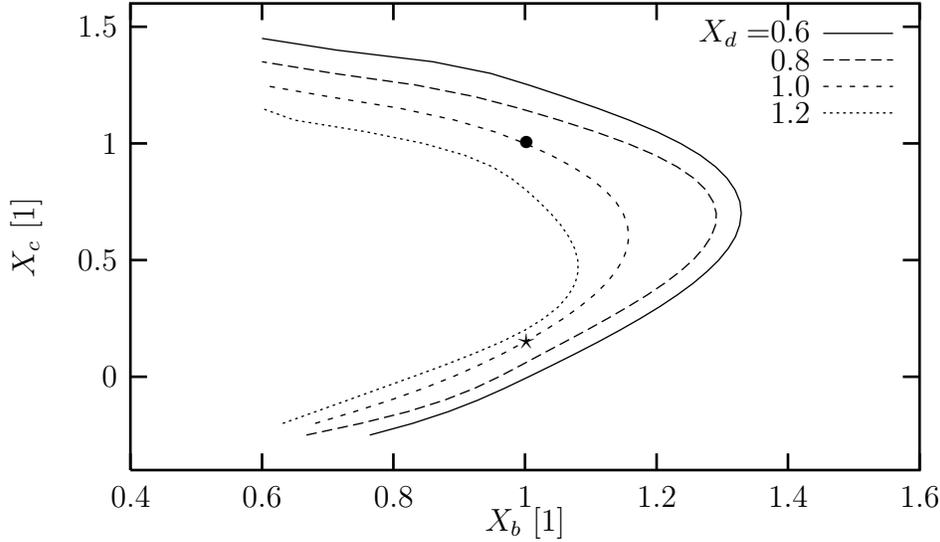}
\caption{Contour plot of the parameters $b$, $c$, $d$ for constant  binding energy. 
}
\label{fig2}
\end{figure}
 Fig. ~2 shows contour plots for different $X_c$ and $X_b$ while keeping $X_d$ at different fixed values. 
Each line in Fig.~2 corresponds to the same binding energy (8.48MeV).  
The black spot indicates the position for  the  original values of the parameters ($X_b$=$X_c$=$X_d$=1).
The star is as in Fig. 1. 
We see that these two solutions for $c$ found above lie on the line in Fig.~2. 
The  $b$- and $c$- values to the left (right) of a  curve for a particular value of $d$
 lead to over-binding (under-binding). 

A crucial rule\cite{Fujita} corresponding to the properties of the $\Delta$ particle 
excitation  mechanism is that the ratio $ b/d $ is 4. 
The Urbana-Argonne\cite{Urbana} 3NF follows this rule, since except for a
phenomenological short range term it includes only the $2\pi$-exchange with an
intermediate $\Delta$ isobar, the Fujita-Miyazawa 3NF.
The TM value for the ratio $b/d$ is 3.43, since in this model the $b$- and
$d$-term include other processes on top of the $2\pi$-exchange with an
intermediate $\Delta$. Also $b$ and $d$ are larger for the TM 3NF than for
the pure $\Delta-2\pi$-exchange. The values for $b$ and $d$ of the Brazil and 
RuhrPot 3NF are close to those of the TM 3NF.
The Texas 3NF, based on chiral perturbation theory, has even larger values
for $b$ and $d$. This shows that it is not at all clear which values for the
strength parameters within the 3NF one should choose.

In Fig.~3 we show the curve for $b$=4$d$ and of course the additional requirement 
that the triton binding energy has the experimental value. As in Fig.~2 the 
underlying NN potential is CD-Bonn. Of course, in looking to Fig.~3 one should 
keep in mind that the choice for the $NN\pi$ form-factor (4) leads to a very
strong dependence of the strength of the 3NF on the cut-off parameter $\Lambda$,
as well.

From Table II in \cite{Friar99} the locations of the $b$ and $c$ parameters for 
several 3NFs are indicated. Note however those 3NFs are not adjusted to the 
triton binding energy together with the CD-Bonn potential.

\begin{figure}[hbt]
\input{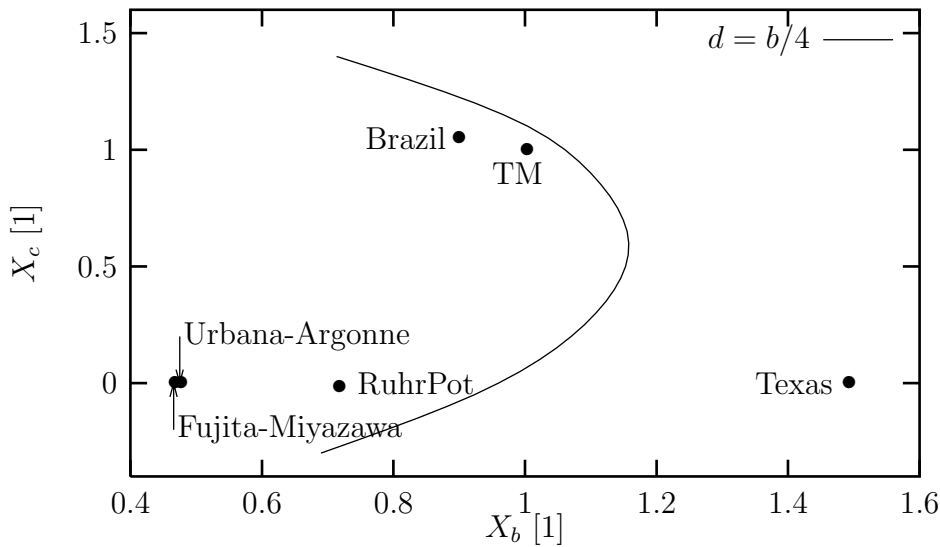}
\caption{Parameters for various 3NFs.}
\label{fig3}
\end{figure}

\section{Summary and outlook}

Stimulated by \cite{Friar99} we studied the binding energy of $^3 $H as a function of the strength parameters 
($a$, $b$, $c$ and $d$ in (\ref{operator}) ) in the TM force. 
We find that the $a$-term is not decisive when varying in the interval 
$ 0 \le a \le 2 $.  The $b$- and $d$-terms, however are  very important to obtain the  
experimental value 8.48MeV. 
Varying $c$ from 0 to 1.5 $\times c_{TM}$ we find that there are two solutions which belong to 
the same binding energy. The new solution is 15\% of the original value, namely,  
0.15 [$m_\pi^{-3}$]. It supports phenomenologically  
the recommendation given in\cite{Friar99} ( based on arguments from chiral perturbation theory) 
that  the short-range part of the 
$c$-term in the TM force should be dropped. 

If one assumes a purely phenomenological point of view for choosing the values 
for $a$ to $d$ in a 3NF of the form (\ref{operator}) Fig.~ 2  
provides a complete overview for the possible values under the requirement 
that the triton binding energy is gained together with the CD-Bonn potential.
Clearly corresponding pictures could be gained for other NN potentials. 
Of course other 3NF mechanisms have to be explored, too. At least for the
$A_y$-puzzle it is clear that the $2\pi$-exchange 3NF is not sufficient to
explain this discrepancy between theory and data. A study of pion range -
short range 3NF terms is underway\cite{Hueber2} where are predicted by 
chiral perturbation theory.

Based on the chosen form (2) and the requirement to fit the triton binding energy
those 3NFs can now be tested in the 3N continuum with high energy.    
At IUCF\cite{IUCF}, RIKEN\cite{Sakamoto,Sakai} and KVI 
measurements are underway  
for 3N observables between 100-300 MeV. These are cross sections and various 
spin observables. They will be analysed using the 3NFs fixed in Fig.~2. 
This might allow to find a preference for a certain region in that
parameter space or will show that additional forms are needed. 
\smallskip


{\bf Acknowlegements}

This paper is  dedicated to  Prof. Dr. Walter Gl\"ockle 
on the occasion of his 60th birthday. 
This work is financially supported by  the Deutsche
Forschungsgemeinschaft under Project No. Gl 87/19-2, No. Hu 746/1-3 and No. 
Gl 87/27-1,
and partially the auspices of the U.S.\ Department of Energy.
The calculations have been performed on the CRAY T3E
of the John von Neumann Institute for Computing, J\"ulich,
Germany.



\begin{thebibliography}{9}




\bibitem{Gloeckle96} Gl\"ockle, W., Wita\l a, H., H\"uber, D., Kamada, H., Golak, J. : 
Phys. Rep. {\bf 274}, 107 (1996). 



\bibitem{Fujita}
Fujita,J.,-I., Miyazawa, H., : Prog. Theor. Phys. {\bf 17}, 360 (1957). 


\bibitem{Tuscon} Coon, S.,A., Schadron, M., D., McNamee, P., C., Barrett, B., R., 
Blatt, D., W., E., McKellar, B., H., J.,: Nucl. Phys. {\bf A 317}, 242 (1979);
Coon, S., A.,: Few-Body Syst., Suppl. {\bf 1}, 41 (1984); 
Coon, S., A., Friar, J., L., : Phys. Rev. {\bf C 34}, 1060 (1986); 
Coon, S., A., Pe\~na, M., T., : Phys. Rev. {\bf C 48}, 2559 (1993). 

\bibitem{Coon81}
Coon, S., A., Gl\"ockle, W., : Phys. Rev. {\bf C 23}, 1790 (1981). 


\bibitem{Brazil}
Coelho, H.,T., Das, T., K., Robilotta, M.,R.,: Phys. Rev. {\bf C 28},1812 (1983); 
Robilotta, M., R., Coelho, H.,T.,: Nucl. Phys. {\bf A 460},645 (1986); 
Murphy, D.,P., Coon, S., A., :Few-Body Syst. {\bf 18}, 73 (1995).

\bibitem{TEXAS}
Ord\'o\~nez, C., Kolck, U., van,: Phys. Lett. {\bf B 291}, 459 (1992); 
Kolck, U., van,: Phys. Rev. {\bf C 49}, 2932 (1994). 


\bibitem{Ruhr}
Eden, J., A., Gari, M., F., : Phys. Rev. {\bf C 53}, 1510 (1996).


\bibitem{Urbana}
Pudliner, B.,S., Pandharipande, V., R., Carlson, J., Pieper, S., C.,
 Wiringa, R., B., : Phys. Rev. {\bf C 56}, 1720 (1970); 
Carlson, J., Pandharipande, V., R., Wiringa, R. B. : Nucl. Phys. {\bf A 401}, 59 (1983).  

\bibitem{cdbonn}
Machleidt, R., Sammarruca, F., Song, Y., : Phys. Rev. {\bf C 53}, 1483 (1996).

\bibitem{av18}
Sperisen, F., Gr\"uebler, W., K\"onig, V., Schmelzbach, P., A., Elsener, K., 
Jenny, B., Schweizer, C., Ulcricht, J.,
 Doleschall, P., : Nucl. Phys., {\bf A422}, 81 (1984).

\bibitem{nijm}
Stoks, V., G., J., Klomp, R., A., M., Terheggen, C., P., F., Swart, J., J., de,
 : Phys. Rev. {\bf C 49}, 2950 (1994).


\bibitem{Koike}
Koike, Y., Haidenbauer, J., : Nucl. Phys. {\bf A 463}, 365c (1987).

\bibitem{Witala3}
Wita\l a, H., H\"uber, D., Gl\"ockle, W., : Phys. Rev. {\bf C 49}, R14 (1994). 

\bibitem{Witala}
Tornow, W., Wita\l a, H., Kievsky, A., : Phys. Rev. {\bf C 57}, 555 (1998).


\bibitem{Hueber}
H\"uber, D., Friar, J., L., : Phys. Rev. {\bf C 58}, 674 (1998).

\bibitem{Kievsky}
Kievsky, A., Viviani, M., Rosati, S., H\"uber, D., Gl\"ockle, W., Kamada, H., Wita\l a, H., Golak, J., : Phys. Rev. 
{\bf C 58}, 3085 (1998).


\bibitem{Sagara}
 Sagara, K.,  Oguri, H.,  Shimizu, S.,  Maeda, K.,  Nakamura, H.,  Nakashima,  T., Morinobu, S., : Phys. Rev. 
{\bf C 50},  576 (1994).  

\bibitem{Koike2}
Koike, Y., Ishikawa, S.,: Nucl. Phys. {\bf A 631}, 683c (1998).

\bibitem{Witala2}
Wita\l a, H., Gl\"ockle, W., H\"uber, D., Golak, J., Kamada, H., : Phys. Rev. Lett. {\bf 81}, 1183 (1998).

\bibitem{Sakamoto}
Sakamoto, N.,  Okamura, H., Uesaka, T.,  Ishida, S., Otsu,  H., 
 Wakasa, T.,  Satou,  Y., Niizeki,  T.,  Katoh,   K.,  Yamashita, T., 
   Hatanaka, K.,  Koike, Y., Sakai, H.,  :
 Phys. Lett. {\bf B 367}, 60 (1996).

\bibitem{Nemoto}
Nemoto, S., Chmielewski, K., Oryu, S., Sauer, P., U.,  : Phys. Rev. {\bf C 58}, 2599 (1998).

\bibitem{Total}
H.\ Wita\l a, H.\ Kamada, A.\ Nogga, W.\ Gl\"ockle, Ch.\ Elster, D.\ H\"uber,
to be appeared in Phys. Rev. {\bf C59} (1999);
W. P.\ Abfalterer {\it et al.},:  Phys.\ Rev.\ Lett.\  {\bf 81}, 57 (1998).

\bibitem{Diff}
H.\ Wita\l a, W.\ Gl\"ockle, D.\ H\"uber, J.\ Golak, H.\ Kamada, {\it Phys.\ 
Rev.\ Lett.\ } {\bf 81}, 1183 (1998);
H.\ Rohdjess {\it et al.}, : Phys.\ Rev.\ {\bf C57}, 2111 (1998).



\bibitem{piNchiral}
Bernard, V., Kaiser, N., Mei\ss ner, Ulf-G., : Int. J. Mad. Phys. {\bf 
E 4}, 193 (1995).
\bibitem{piNchiral2}
 Fettes, N., Mei\ss ner, Ulf-G., Steiniger, S., : Nucl. Phys. {\bf A 640}, 199(1998).

\bibitem{NNchiral}
Ord\'o\~nez, C., Ray, L., Kolck, U., van, : Phys. Rev. {\bf C 53}, 2086 (1996).

\bibitem{NNchiral2}
Park, T.-S.,  Kubodera, K.,  Min, D.,-P.,   Rho, M., : Phys. Rev. {\bf C 58}, 637 (1998).

\bibitem{NNchiral3}
Epelbaoum, E., Gl\"ockle, W., Mei\ss ner, Ulf-G., : Nucl. Phys. {\bf A 637}, 
107 (1998);  Epelbaoum, E., G\"ockle, W., Kr\"uger, A., Mei\ss ner, Ulf-G.,
: Nucl. Phys. {\bf A 645}, 413 (1999).

\bibitem{Friar99} Friar, J., L., H\"uber, D., Klock, U., van,: Phys. Rev. {\bf C59}, 53, (1999).



\bibitem{Hueber2}
H\"uber, D., private communication.



\bibitem{Stadler}
Stadler, A., Gl\"ockle, W., Sauer, P., U., : 
Phys. Rev. {\bf C 44}, 2319 (1991);  
Stadler, A.,  Adam, J., Jr., Henning, H.,   Sauer, P., U., :
Phys. Rev. {\bf C 51}, 2896 (1995). 

\bibitem{Nogga}
Nogga, A., H\"uber, D., Kamada, H., Gl\"ockle, W. :
Phys. Lett. {\bf B 409}, 19 (1997).

\bibitem{Hempen98}
Hempen, P., et al., : Phys. Rev. {\bf C57}, 484 (1998).



\bibitem{IUCF}
B.D. Anderson, B., D.,  et al., : Nucl. Phys. {\bf  A 631}, 752c  (1998). 

\bibitem{Sakai} 
Sakai, H., Sekiguchi, K., : private communication.

\end{thebibliography}
\end{document}